# Dramatic enhancement of second and third harmonic generation in gold nanogratings in the visible and UV ranges

S. Mukhopadhyay[1], L. Rodriguez-Suné[1], C. Cojocaru[1], K. Hallman[2], G. Leo[3], M.A. Vincenti[4], M. Belchovski[4], D. de Ceglia[4], M. Scalora[5], J. Trull[1]

[1]*Department of Physics, Universitat Politècnica de Catalunya, Rambla Sant Nebridi 22, 08222 Terrassa (Barcelona), Spain*
[2] *PeopleTec, Inc. 4901-I Corporate Dr., Huntsville, AL 35805, USA*
[3]*Laboratoire Matériaux et Phénomènes Quantiques, Université Paris Cité & CNRS, 10 rue Alice Domon et Léonie Duquet, 75013 Paris, France*
[4]*Department of Information Engineering – University of Brescia, Via Branze 38, 25123 Brescia, Italy*
[5] *Aviation and Missile Center, US Army CCDC, Redstone Arsenal, AL 35898-5000, USA*

**Abstract**

Notwithstanding its long history, the study of nonlinear optics from metal surfaces is still an active field of research. For instance, in view of the presence of absorption questions remain concerning the possibility of significantly enhancing harmonic conversion efficiencies in the visible and UV ranges. While to many it may seem that metals do not easily lend themselves to that purpose, they are nevertheless crucial materials in the development of nanophotonics, and more generally, to electromagnetism at the nanoscale. Here, we report our experimental observations and numerical simulations of second and third harmonic generation from a gold nanograting, which exhibits a plasmonic resonance whose spectral position depends on incident angle. All things being equal, the enhancement of nonlinear optical processes from the UV to the near IR range manifests itself in dramatic manner: second harmonic generation conversion efficiencies increase more than three orders of magnitude compared to a flat gold mirror, while third harmonic generation conversion efficiency increases by nearly four orders of magnitude, both in excellent agreement with predictions. The clear inferences one may draw from our results are that our model describes the dynamics with unprecedented accuracy, and that much remains to be revealed in the development of nonlinear optics of metals at the nanoscale.

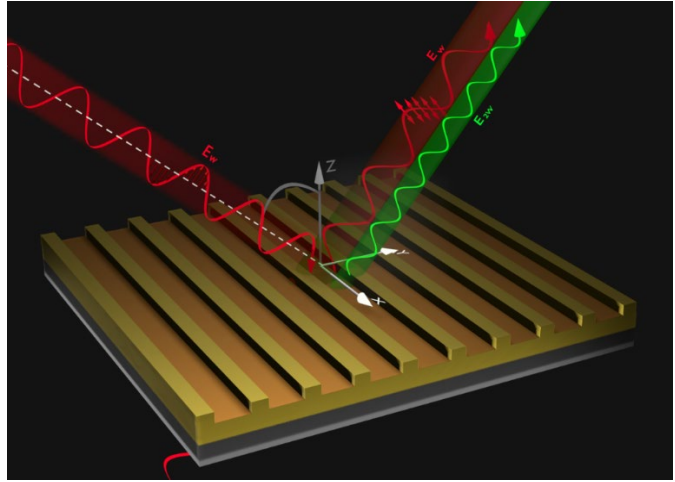

**Introduction**

Nonlinear frequency conversion in metals has been the subject of theoretical and experimental research since the emergence of nonlinear optics, partly motivated by their high third order nonlinearities despite their high absorption in the visible and near infrared ranges. The works by Jha [1], Brown et al., [2,3], Bloembergen et al., [4] and others discussed the various contributions of free electrons to second harmonic generation (SHG) in silver and, to a far more limited extent, the role played by bound electrons, while emphasizing the fact that it is possible to obtain a nonlinear, second order response from media with inversion symmetry. Other contributions later explored SHG in metals based on the hydrodynamic model [5], as well as studies of surface-enhanced SHG in silver gratings [6], always with emphasis on the dynamics of free electrons only. The absence of a phase-matching mechanism in surface-enhanced SHG severely limits conversion efficiencies, while the presence of absorption limits penetration depth into the metal, thus inhibiting higher order bulk nonlinearities and rendering these materials inefficient frequency converters.

The interest in metals' potential and usefulness as optical materials, beyond their use as mirrors, was renewed following the inception of the field of photonic crystals [7-9]. Within the last few years, the study of nonlinear frequency conversion at the nanoscale has attracted a great deal of attention, while developing along several main lines: plasmonics [10], dielectric metasurfaces [11], nanowire arrays [12], or combined systems of resonances that exploit metal-dielectric nano-antenna resonances, together with transparent conductive oxides like ITO that display epsilon-near-zero (ENZ) features [13-15]. These approaches exploit geometrical features, like Fano, plasmonic, and so-called magnetic resonances, and/or intrinsic material singularities like ENZ conditions, that tend to enhance the local fields near metal points and edges, inside nanowire or antenna assemblies and resonant metasurfaces [16].

Metal nanostructures can enhance optical responses by means of surface-enhanced spectroscopies such as Raman scattering [17], infrared absorption and fluorescence spectroscopies [18]. They can also generally improve the efficiency of surface phenomena, and for the preparation of specimen materials, depending on the purpose of the measurement. These methods can achieve the sensitivity of single molecule level and provide unique insights on molecular behaviors and especially on surface properties in many research fields [19].

Current research in nano-structured materials and linear plasmonic phenomena like subwavelength resolution [20], enhanced transmission [21] or absorption, has directed renewed attention on the origin of harmonic generation in sub-wavelength, modulated materials [22]. Recent theoretical models have predicted the enhancement of absorption and harmonic generation when a nonlinear medium is placed near a metal grating composed of rectangular slits. This structure supports wide-band Fabry-Perot-like resonances [23], whose dispersion is significantly altered by the presence of the grating. The main effect of the periodicity is the activation of surface modes [24] that may enhance second and/or third harmonic generation even from materials with relatively small nonlinear coefficients. Large enhancement of a four-wave mixing signal in plasmonic nanocavity gratings was demonstrated at normal incidence [25].

We recently carried out a combined theoretical and experimental study of second and third-harmonic generation in thin-layered gold surfaces, with the purpose of extracting the basic physical intrinsic properties of the material. We used an expanded version of the hydrodynamic model to account for the linear and nonlinear material dispersions associated with both free and bound electrons, including hot electrons [26]. Using those parameters, in what follows we report the experimental observation of the enhancement of second and third harmonic signals in a gold grating by three and nearly four orders of magnitude, respectively, relative to a flat metal layer, in excellent agreement with our simulations

**Sample parameters and linear characterization**

The grating consists of periodic parallel nanogrooves etched on the surface of a thick gold layer substrate, having a period $p$, channel width $a$, and depth $w$, as schematically shown in Figure 1(a). We fabricated such a grating by first depositing a 200nm-thick gold layer on top of a $SiO_2$ substrate. Subsequently we performed electron-beam lithography on a PMMA resist and deposited an additional thickness of gold that would ultimately form the grooves. Finally, we removed the resist with acetone. A SEM picture of one of the fabricated nanogratings is shown in Figure 1(b). The period is $p≈610$nm, and channel width $a≈385$nm, with an uncertainty less than 10nm.

The theoretical spectral response of the grating in reflection is calculated by means of rigorous coupled wave analysis (RCWA) and is shown as a function of angle in Figure 1(c). The resonance is seen to shift from 700 nm up to 1100 nm when the angle of the incident beam varies from 0º up to 50º. For example, for an incident angle of ~16º the resonance is located near 800 nm; at ~37º the resonance shifts to 1000 nm; at ~50º the resonance appears at 1100nm. We performed the linear characterization of the grating using a supercontinuum laser source (Fyla SCT 1000) by measuring the reflected spectrum as a function of incidence angle with a high-resolution spectrometer.

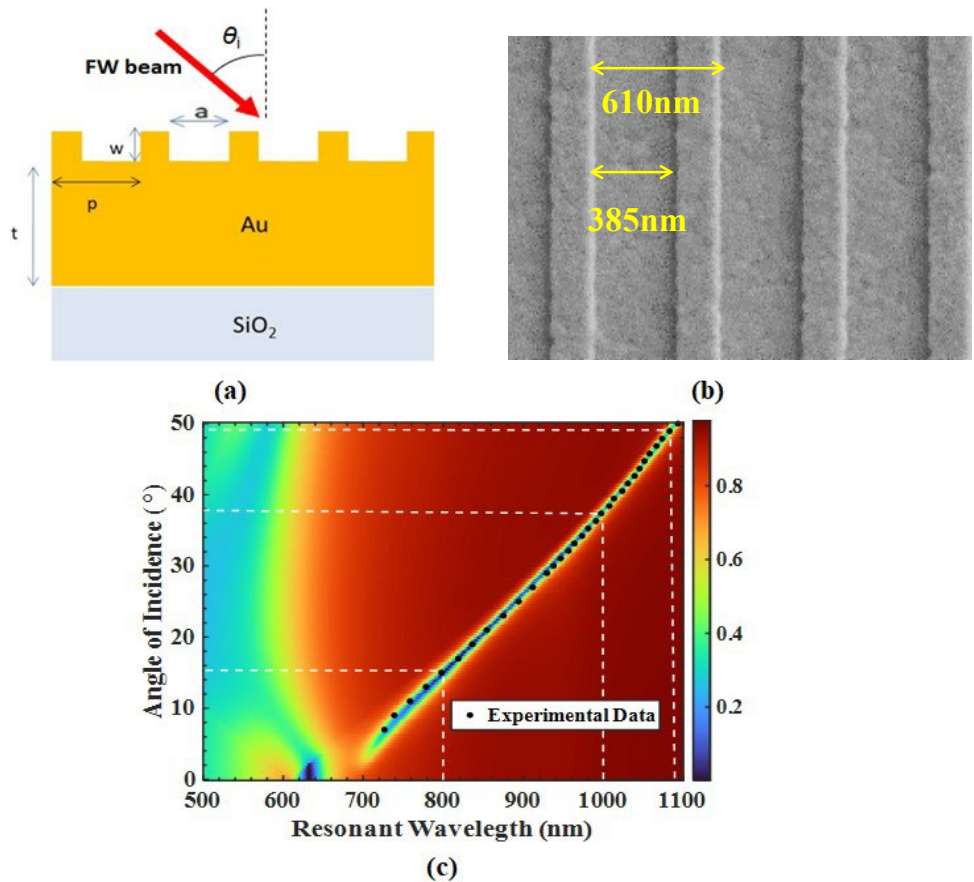

**Figure 1** (a) Schematic representation of the gold nanograting for oblique incidence. (b) SEM image of one of the fabricated gratings; (c) Simulation of the grating reflectance as a function of incident angle and wavelength for the grating having a=385 nm, w=62 nm, p=605 nm and t = 200 nm. Experimental measurements of the resonance wavelength as a function of angle are shown as black dots.

The central wavelengths of the resonance measured for incident angles ranging from 7º to 50º and TM-polarized incident light are displayed as dots in Figure 1c, in excellent agreement with our simulations. As expected, no resonances appear when the grating is illuminated with TE-polarized light. Figure 2(a) shows an example of a resonance located near 800 nm (simulation and experiment). The reflectance is normalized to the value obtained from the unpatterned gold layer. In Figure 2(b) we show the predicted electric and magnetic field distributions around one of the grooves. The plasmonic structure induces large local fields near the surface of the metal, leaving the evanescent tail of the field to harvest the effects of both surface and bulk nonlinearities. The fields behave similarly when tuned near 1000nm or 1100nm, although the increased value of the dielectric constant increases the effective potential the wave experiences, thus enhancing the importance of surface phenomena at the expense of the bulk response. At resonance, the plasmonic grating confines the fields near the surface, producing an intense field close to the corners. A bound wave travels along the grating, with an evanescent tail that spills just inside the metal [Figure 2(b)], which in turn absorbs all the incident energy resulting in near-zero reflections. The presence of the intense field at the edges of the grating is responsible for pump absorption and the simultaneous enhancement of nonlinear interactions.

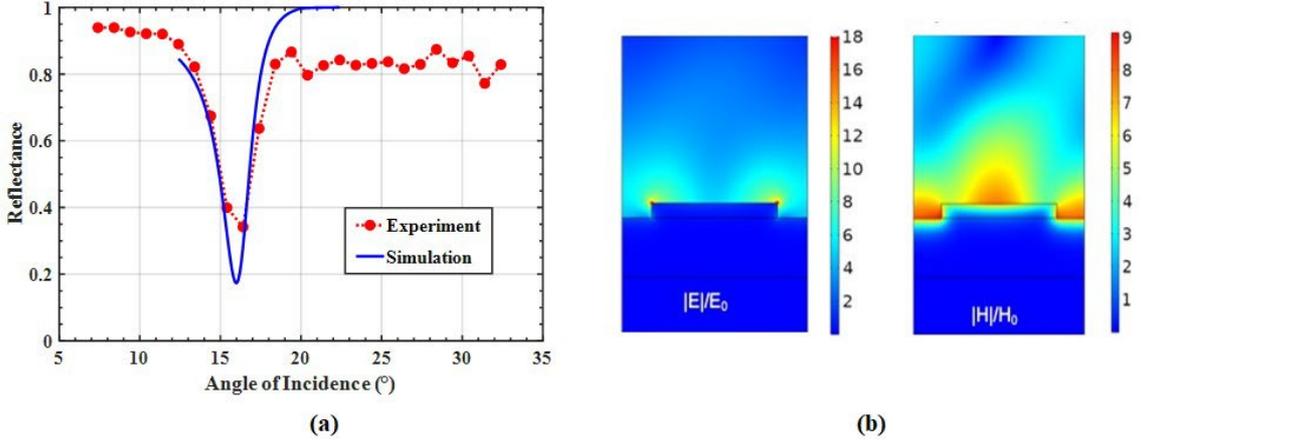

**Figure 2** (a) Reflectance as a function of incident angle for carrier wavelength tuned at 800nm, normalized to the value obtained from the unpatterned gold layer: simulation (solid blue curve) and experimental measurements (dashed curve with red dots). The resonance is centered at 16º (b) Electric and magnetic field distributions around a single groove, normalized with respect to the incident electric and magnetic field amplitudes.

**Brief outline of the model.**

We simulated nonlinear light-matter interactions using a theoretical model described elsewhere [21,27-29], and modified here accordingly to address the wavelength range under consideration. The classical description of free and bound electron current densities, defined as $\mathbf{J}_f = \dot{\mathbf{P}}_f$, and $\mathbf{J}_{bj} = \dot{\mathbf{P}}_{bj}$, respectively, leads to the following scaled equations of motion for conduction and core electrons:

$$\ddot{\mathbf{P}}_f + \tilde{\gamma}_f \dot{\mathbf{P}}_f = \frac{e^2 \lambda_0^2 n_{0,f}}{m_0^* c^2}\mathbf{E} + \tilde{\Lambda}(\mathbf{E}\bullet\mathbf{E})\mathbf{E} - \frac{e\,\lambda_0}{m_0^* c^2}(\nabla\bullet\mathbf{P}_f)\mathbf{E} + \frac{e\,\lambda_0}{m_0^* c^2}\dot{\mathbf{P}}_f \times \mathbf{H} \\ + \frac{3E_F}{5m_0^* c^2}\left(\nabla(\nabla\bullet\mathbf{P}_f) + \frac{1}{2}\nabla^2\mathbf{P}_f\right) - \frac{1}{n_{0,f}e\lambda_0}\left[(\nabla\bullet\dot{\mathbf{P}}_f)\dot{\mathbf{P}}_f + (\dot{\mathbf{P}}_f\bullet\nabla)\dot{\mathbf{P}}_f\right] \quad (1)$$

$$\ddot{\mathbf{P}}_{bj} + \tilde{\gamma}_{bj}\dot{\mathbf{P}}_{bj} + \tilde{\omega}_{0,bj}^2 \mathbf{P}_{bj} - \tilde{\beta}_j(\mathbf{P}_{bj}\bullet\mathbf{P}_{bj})\mathbf{P}_{bj} + \tilde{\Theta}_j(\mathbf{P}_{bj}\bullet\mathbf{P}_{bj})^2\mathbf{P}_{bj} - \tilde{\Delta}_j(\mathbf{P}_{bj}\bullet\mathbf{P}_{bj})^3\mathbf{P}_{bj} = \\ \frac{n_{0,b}e^2\lambda_0^2}{m_{bj}^* c^2}\mathbf{E} + \frac{e\,\lambda_0}{m_{bj}^* c^2}(\mathbf{P}_{bj}\bullet\nabla)\mathbf{E} + \frac{e\,\lambda_0}{m_{bj}^* c^2}\dot{\mathbf{P}}_{bj}\times\mathbf{H} \quad (2)$$

The subscript *b* stands for *bound*, and the counter *j* represents the *j*th bound oscillator species. The linear dielectric response of gold is modelled using one Drude component and three separate Lorentzian functions to represent the dielectric response more accurately in the 200-300nm range.

Eqs.(2) thus represent three separate equations each describing an oscillator species. The local dielectric constant may be written as:

$$\varepsilon(\omega) = 1 - \frac{\tilde{\omega}_{pf}^2}{\omega^2 + i\tilde{\gamma}_f \omega} - \frac{\tilde{\omega}_{p1}^2}{\omega^2 - \tilde{\omega}_{01}^2 + i\tilde{\gamma}_{01}\omega} - \frac{\tilde{\omega}_{p2}^2}{\omega^2 - \tilde{\omega}_{02}^2 + i\tilde{\gamma}_{02}\omega} - \frac{\tilde{\omega}_{p3}^2}{\omega^2 - \tilde{\omega}_{03}^2 + i\tilde{\gamma}_{03}\omega}, \qquad (3)$$

where $(\tilde{\omega}_{pf}, \tilde{\omega}_{p1}, \tilde{\omega}_{p2}, \tilde{\omega}_{p3}) = (7.1, 3.4, 4.79, 6.35)$, $(\tilde{\gamma}_f, \tilde{\gamma}_{01}, \tilde{\gamma}_{02}, \tilde{\gamma}_{03}) = (0.05, 1.25, 1.45, 1.25)$, and $(\tilde{\omega}_{0f}, \tilde{\omega}_{01}, \tilde{\omega}_{02}, \tilde{\omega}_{03}) = (0, 2.45, 3.45, 4.75)$. Eq.(3) is obtained by Fourier transforming Eqs.(1-2) after dropping all nonlinear and nonlocal terms. The frequency $\omega = 1/\lambda$, where $\lambda$ is in microns. Eqs.(1-2) amount to an expanded hydrodynamic model that includes a description of surface ($(\nabla \bullet \mathbf{P}_f)\mathbf{E}$), magnetic ($\dot{\mathbf{P}}_f \times \mathbf{H}$), convective $\left[(\nabla \bullet \dot{\mathbf{P}}_f)\dot{\mathbf{P}}_f + (\dot{\mathbf{P}}_f \bullet \nabla)\dot{\mathbf{P}}_f\right]$, nonlocal $\left(\nabla(\nabla \bullet \mathbf{P}_f) + \frac{1}{2}\nabla^2 \mathbf{P}_f\right)$ and hot electron ($\tilde{\Lambda}(\mathbf{E} \bullet \mathbf{E})\mathbf{E}$) nonlinearities in the free electron polarization component. Furthermore, Eqs.(2) describe surface ($(\mathbf{P}_{bj} \bullet \nabla)\mathbf{E}$), magnetic ($\dot{\mathbf{P}}_{bj} \times \mathbf{H}$) and bulk ($-\tilde{\beta}_j(\mathbf{P}_{bj} \bullet \mathbf{P}_{bj})\mathbf{P}_{bj} + \tilde{\Theta}_j(\mathbf{P}_{bj} \bullet \mathbf{P}_{bj})^2 \mathbf{P}_{bj} - \tilde{\Delta}_j(\mathbf{P}_{bj} \bullet \mathbf{P}_{bj})^3 \mathbf{P}_{bj}$) nonlinearities triggered by bound electrons. We neglect even order nonlinearities in centrosymmetric systems. We choose to expand these bound electron bulk terms up to seventh order because the local fields can be quite large near edges and points. For instance, Figure 2(b) suggests a local electric field *intensity* enhancement by a factor of nearly 400. As a result, just a few GW/cm$^2$ incident onto the sample can turn into several TW/cm$^2$, thus necessitating going beyond a mere third order response. In fact, the higher order terms shown also contribute to odd harmonics. For instance, a few of the many non-instantaneous terms contributing to the full, complex $\chi_\omega^{(3)}$ response are proportional to $\tilde{\beta}_j |P_\omega|^2 P_\omega$, $\tilde{\Theta}_j |P_\omega|^4 P_\omega$, and $\tilde{\Delta}_j |P_\omega|^6 P_\omega$, where $\tilde{\beta}_j = \omega_{0,j}^2 \lambda_0^2 / (L^2 n_{0b}^2 e^2 c^2)$, and $\tilde{\Theta}_j = \frac{\tilde{\beta}_j}{n_{0b}^2 e^2 L^2}$ $\tilde{\Delta}_j = \frac{\tilde{\beta}_j}{n_{0b}^4 e^4 L^4}$. Similarly, some of the terms contributing to the full, complex $\chi_{3\omega}^{(3)}$ are proportional to $\tilde{\beta}_j P_\omega^3$, $\tilde{\Theta}_j |P_\omega|^2 P_\omega^3$, $\tilde{\Delta}_j |P_\omega|^4 P_\omega^3$. It should be evident that the approach outlined above may be considered more accurate than the usual, instantaneous responses of the type $\chi^{(3)} |E_\omega|^2 E_\omega$ and $\chi_{3\omega}^{(3)} E_\omega^3$, which require prior knowledge of nonlinear dispersion. On the other hand, $\tilde{\beta}_j$, $\tilde{\Theta}_j$, and $\tilde{\Delta}_j$ are real, known, and depend on linear oscillator parameters like resonance frequency $\omega_{0,j}$ and lattice constant $L$. These considerations about free and bound electron third and higher order nonlinearities are important because they can cause a dynamic shift of the plasmonic resonance, either towards the red or the blue, depending on tuning relative to the bound electron resonances [29]. In Eqs. (1) and (2), $n_{0,f}$ and $n_{0,b}$ are the free and bound electron densities, respectively, in the absence of an applied field; the subscript *j* in Eqs.(2) indicates the presence of multiple bound electron species, three in our case, for simplicity assumed to have the same particle density $n_{0,b}$; *e* is the electronic charge; $E_f$ is the Fermi energy; $\lambda_0 = 1\mu m$ is a conveniently chosen scaling wavelength; *c* is the speed of light in vacuum; $m_0^* = m_{bj}^* = m_e$ are the free and bound electron effective masses, for simplicity assumed to be equal to the free electron mass; $\xi = z/\lambda_0, \varsigma = y/\lambda_0, \zeta = x/\lambda_0, \tau = ct/\lambda_0$ are the scaled space and time coordinates that we used to calculate spatial and temporal derivatives. Finally, plasma and resonance frequencies and damping coefficients are scaled as follows: $\tilde{\omega}_p^2 = \frac{4\pi n e^2}{m^*}\frac{\lambda_0^2}{c^2}$, $\tilde{\omega}_0 = \omega_0 \frac{\lambda_0}{c}$, $\tilde{\gamma} = \gamma \frac{\lambda_0}{c}$. For additional, specific details about the model and method of solution, we refer the reader to references [21,27-29]. Here we emphasize that in the wavelength range below 1 μm it is *essential* to model the metal as a collection of free and bound charges as the incident wave begins to probe atomic orbitals that reside below the free electron cloud. The integrations are carried out using a fast Fourier transform beam propagation method using a spatial step δz = δy = δx = 1.25 nm, and a correspondingly causal (δz = cδt) temporal integration step δt = 4×10⁻¹⁸ sec. However, using δz = δy = δx = 2.5 nm and a correspondingly larger time step yields similar results. We note that unlike unconditionally stable spectral methods, ordinary finite difference time domain (FDTD) methods are subject to numerical stability criteria that in a multidimensional grid necessarily disrupt the causality condition, leaving open the possibility of introducing phase errors and numerical dispersion [30], especially in highly nonlinear environments. To our knowledge, these issues and concerns remain unaddressed.

**Experimental setup**

The setup for the nonlinear characterization of the nanograting is shown schematically in Figure 3. We perform a first set of measurements using a Ti:Sapphire femtosecond laser tunable around 800 nm, emitting pulses of 170 fs (FWHM in intensity) at 76 MHz repetition rate, with a CW output average power of 1 W. Intensities between 1 and 4 GW/cm² were measured when a lens of either 20 cm or 10 cm focal length was used to focus the beam on the sample plane. A half-wave plate, placed at the laser output, controls the polarization of the fundamental beam. The grating is set in a rotary support that controls the incident angle and placed between two different filters to avoid possible harmonic signals arising from portions of the setup far from the sample. Before the sample we have a long pass filter that eliminates any background SH from the source. Right after the sample a bandpass filter centered around the SH wavelength attenuates the fundamental field that has passed through the sample. This avoids possible harmonic signals arising from portions of the setup far from the sample. However, this filter is not enough to eliminate all the fundamental radiation, and we place another bandpass filter after the collimating lens and two spectral filters having 20nm bandpass transmission around the SH frequency, one of which is placed just in front of the detector. By choosing these filters, the total optical density exceeded the harmonic generation efficiency, making sure that only SH radiation arrives at the detector. A Wollaston polarizer separates the TM and TE polarization components of the SH signal, allowing simultaneous measurements of both polarizations. We expect values of SH efficiency of order of $10^{-9}$ or $10^{-10}$, requiring a very sensitive detection system. Light is collected with a photomultiplier (PMT) where the detected

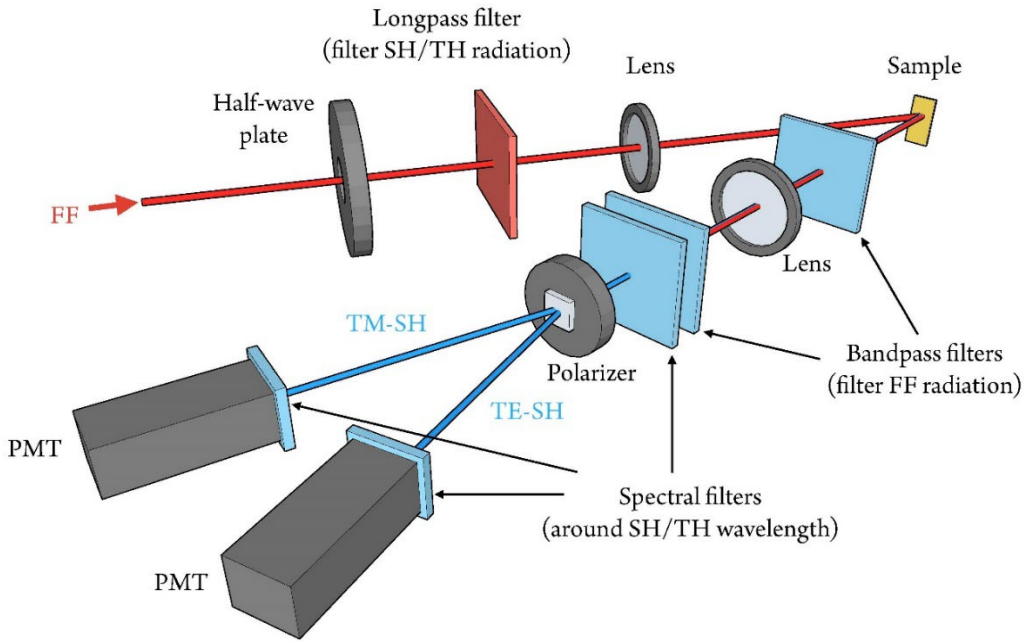

**Figure 3:** (a) Sketch of the experimental set-up used to measure reflected second harmonic signals generated by the gold grating as a function of angle of incidence, polarization and incident wavelength.

Signal is strongly amplified. The entire detection system is mounted on a rigid platform, which in turn is placed on a rotating tail to allow measurement of either transmittance or reflectance.

A second set of measurements was performed using an amplified Ti:Sapphire laser system pumping an optical parametric amplifier, from which tunable 100 fs pulses at 1 kHz repetition rate were obtained. In this experiment, we used incident wavelengths of 1000 nm and 1100 nm, respectively. With a set of two polarizers, the energy *per* pulse was controlled, fixed at 100 nJ when measurements were taken, which led to peak power densities of 1.5 GW/cm$^2$ when a lens of focal length 500 cm was used to focus the beam on the sample plane. The detection system for these experiments was built in similar fashion, but different filters were used, suitable for these fundamental and SH wavelengths. We performed a detailed calibration procedure to estimate the efficiencies of the process as accurately as possible. A strong SH signal was generated with a BBO crystal, with which the PMT responsivities for each wavelength could be measured comparing the signal read by a power meter and the PMT. Then, the efficiencies were calculated as the ratio between SH and fundamental energies. For the detection of the TH signals, we used special optics and filters with enhanced performance in the UV. Commercial UV filters (generally bandpass filters) usually transmit less than 50% of the incident light, and the performance of lenses and mirrors deteriorates. Furthermore, the sensitivity of the PMT is reduced drastically. For these reasons there is some difficulty in estimating experimental TH efficiencies accurately. We limited ourselves to assessing efficiencies based on the transmittance values of the filters and the responsivity of the PMT at the TH wavelength.

**Results and discussion**

In the first experiment the input wavelength is tuned near 800nm, mapping the resonance that appears at 16º incident angle (Figure 2a). Given a grating periodicity of approximately 610nm, the fundamental beam does not generate additional diffraction orders. However, its SH (400nm) and TH (266nm) will produce several diffraction orders. Figures 4a and 4b show the calculated diffraction angles at the second and third harmonic wavelengths, respectively. In order to evaluate the relative contributions of the different diffraction orders we fixed the incident angle to 16º and the wavelength to 800nm and made an angular scan of the emitted SH and TH radiation. For the SH (Figure 4c) we find an intense zeroth order emission at 16º, and first order emission at 63º approximately 4 times weaker than the first order. For the TH (Figure 4d) the first diffraction order is approximately 3 times *more* intense than the zeroth order. The relative amounts of radiation between the different orders of the SH emission are the same as the linear result obtained by illuminating the grating with radiation at 400nm. Limitations in our experimental setup allowed us to measure only the diffraction orders appearing on the side of the reflecting beam.

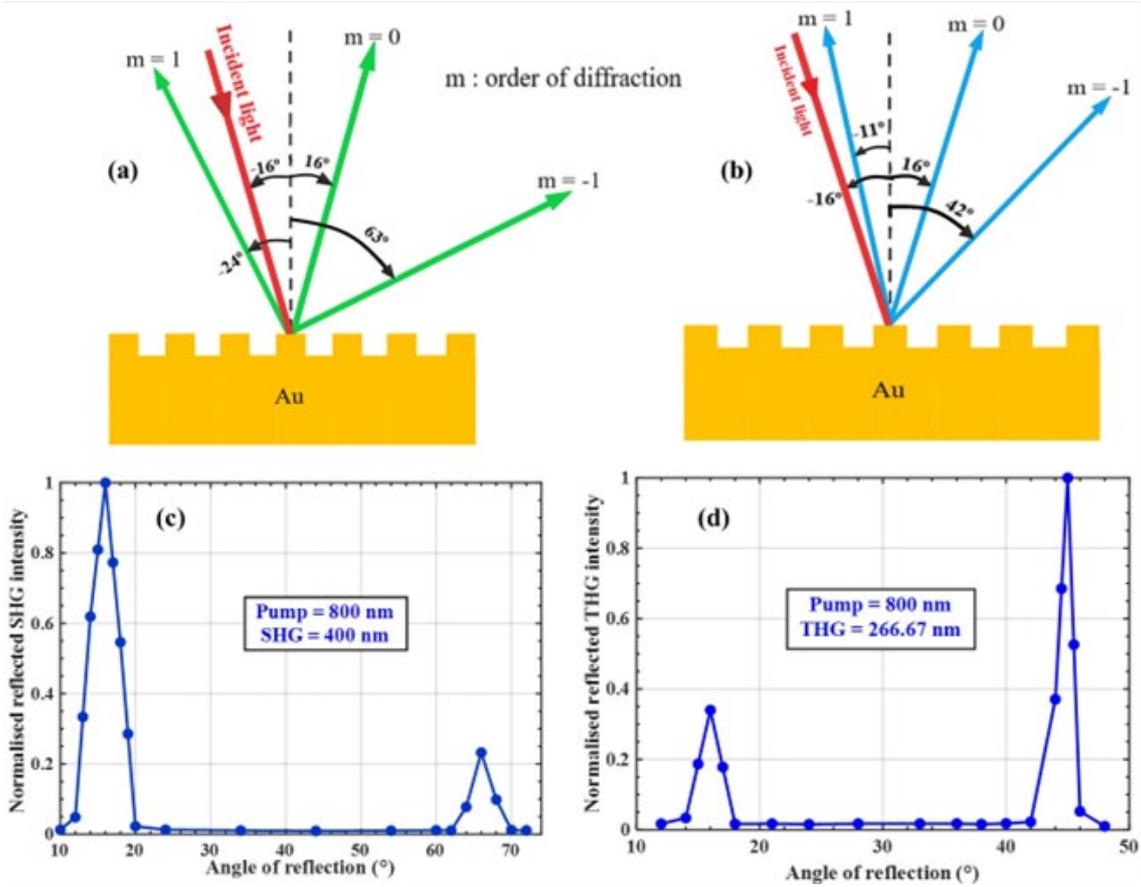

**Figure 4:** (a) Schematic representation of the diffraction orders for the SH (a) and TH (b) wavelengths. Angular measurement of the reflected SH (c) and TH (d) signals.

**Second harmonic generation**

The spectral response of TM-polarized SH emissions is summarized in Figure 5 for TM-polarized pump pulses. The SH response of the surface is triggered mostly by the free electron components. Figure 5a shows the measurements of the reflected SH efficiency for 16º fundamental incident angle when the carrier wavelength of the incident pulse varies across the resonance. A maximum efficiency of ~$1.1 \times 10^{-9}$ is obtained at 16º, and a maximum efficiency of ~$0.4 \times 10^{-9}$ was measured at the first diffraction order, 63º. One additional order is expected to emanate at the negative angle of -24º, but we detect signals propagating away from the grating only on one side of the incident beam. The total efficiency calculated by adding up the two diffraction orders yields a maximum efficiency of $1.5 \times 10^{-9}$. As reference, the SH efficiency coming from the unpatterned portion of the same sample was measured under identical experimental conditions, yielding an efficiency of $2 \times 10^{-12}$, shown on the right-axis of Figure 5a.

The enhancement factor of the SH signal is calculated by scaling the efficiency that includes the zeroth and first diffraction orders to the reference SH originating from the unpatterned portion of the gold sample. We obtained a maximum enhancement factor of 1400, as shown in Figure 5b. The simulations are carried out using pulses of increasing duration, are normalized with respect the efficiency of bare gold, and are also shown in Figure 5b. The results clearly show excellent agreement between theory and experimental observations, with near-perfect reproduction of shape, width, and maximum amplitude of the spectral response for comparable pulse durations. The only free parameters used in the calculations are the effective masses of free and bound electrons, along with a variable peak power density, which is modulated between 1GW/cm$^2$ and 4GW/cm$^2$.

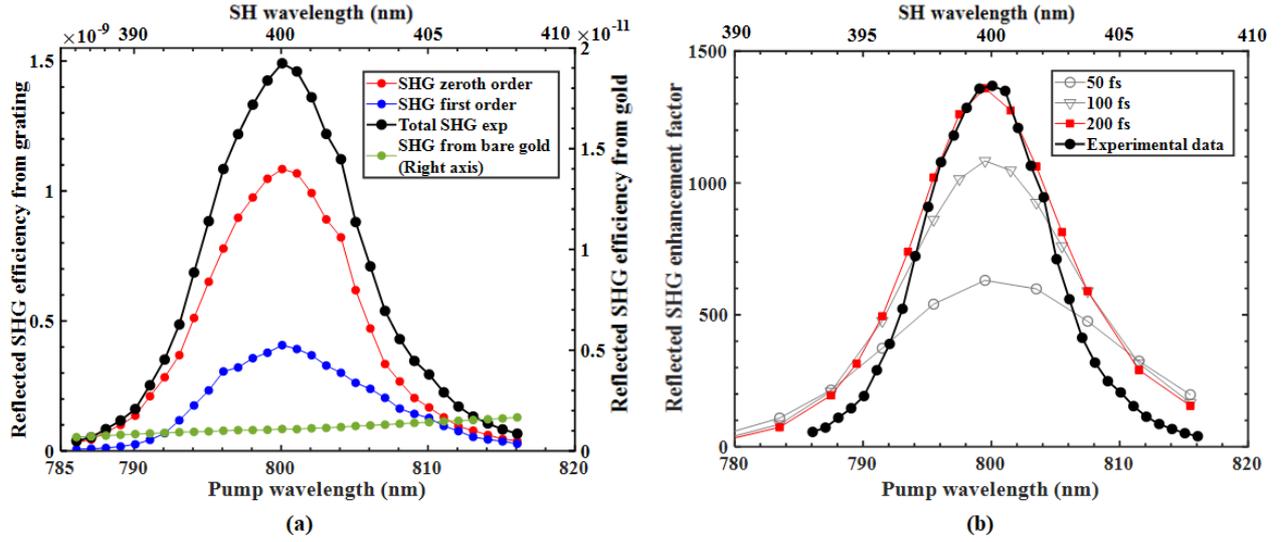

**Figure 5:** a) Measured efficiency of reflected SHG as a function of the carrier wavelength at 16º (blue curve), 63º (red curve), and total (black curve) normalized to the reflected SH efficiency of a bare gold mirror (green curve at the bottom, right axis). (b) Total measured (black curve, same as in 5a) and simulated reflected SHG efficiencies from the grating using pulses having FWHM of different durations, as indicated. The simulation gathers both diffracted orders simultaneously.

Since the resonance wavelength changes by varying the angle of incidence, similar enhancement is expected to occur at different wavelengths by merely tilting the sample. SH efficiency was measured at two alternative incident angles and wavelengths. We measured the reflected SH efficiency at the zeroth diffraction order at 37º incident angle, where the laser carrier wavelength maps the resonance located around 1000nm, and at 50º by tuning the fundamental wavelength around 1100nm. We show the combined results of the three measured wavelengths in figure 6, where we plot the absolute conversion efficiencies vs. incident angle for the three different tuning conditions we explored. The results obtained at these different angles also show three orders of magnitude enhancement, although different incident angles and penetration depths into the metal can lead to different overall conversion efficiencies.

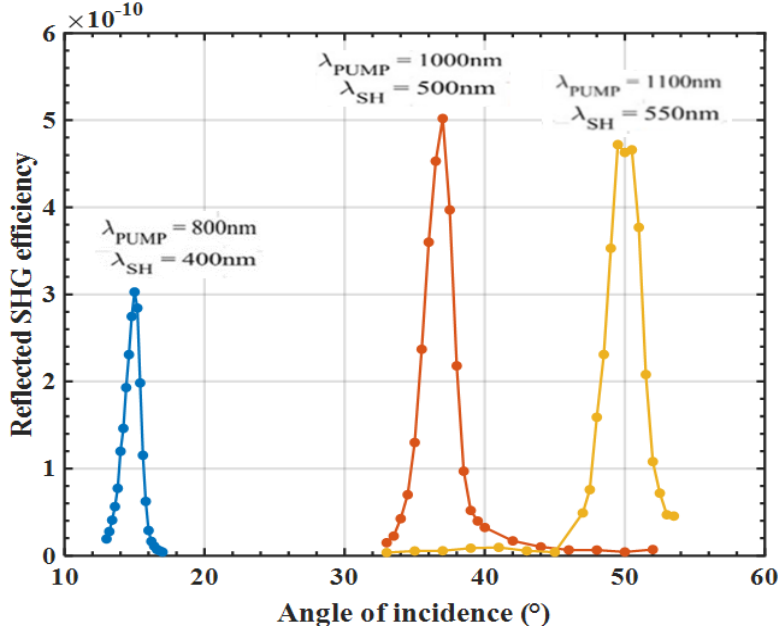

**Figure 6:** Experimental measurements of SHG as a function of angle for incident pulses having different carrier wavelengths. Conversion efficiencies remain of the same order of magnitude across the entire range, although overall efficiencies maxima may vary somewhat due to slightly different effective parameters.

**Third Harmonic Generation**

THG has two possible sources: hot electrons in Eq.(1), and bound electron contributions in Eqs.(2). Tuning the pump at 800nm places it in a regime dominated by the third order nonlinearity of free electrons [26], i.e., the term $\tilde{\Lambda}(\mathbf{E} \bullet \mathbf{E})\mathbf{E}$. In contrast, the third harmonic signal at 266nm is subject to the nonlinear dispersion of bound electrons and interband transitions, driven by the terms $-\tilde{\beta}_j(\mathbf{P}_{bj} \bullet \mathbf{P}_{bj})\mathbf{P}_{bj} + \tilde{\Theta}_j(\mathbf{P}_{bj} \bullet \mathbf{P}_{bj})^2 \mathbf{P}_{bj} - \tilde{\Delta}_j(\mathbf{P}_{bj} \bullet \mathbf{P}_{bj})^3 \mathbf{P}_{bj}$. The spectral response of the TH emission is depicted in Figure 7. The totality of the TH signal emerging only on one side of the incident beam is composed of the zeroth (16°) and the first (44°) diffraction orders, with nearly 70% of the light now exiting at 44°. The experimental data are depicted in Figures 7a and 7b, showing a maximum enhancement factor of nearly 4000.

Just as we did for SHG, here too we use pulses of increasing duration to highlight the dependence of conversion efficiency on the frequency content of the incident pulse. We consider the combined emission from the zeroth and first diffraction orders and normalize conversion efficiencies with respect to the TH efficiency of a bare, unpatterned gold surface. The results of the simulations are shown in Figure 7b and are overlapped with the total TH enhancement factor reported in Figure 7a. Once again, the comparison between experimental result and simulation are in excellent agreement in terms of shape, spectral width, and maximum amplitude. While estimates of experimental TH conversion efficiencies are limited by filter performance and absorption, our simulations suggest they should be of order $10^{-6}$ or better. In addition, it may be possible to harness significantly larger enhancement factors with slight geometrical modifications. This situation is depicted in Figure 8, where we plot the results of simulations using 200fs pulses incident on a grating having the same periodicity p=605nm, and by gradually increasing (decreasing) channel (ridge) width, showing a maximum enhancement of nearly four orders of magnitude.

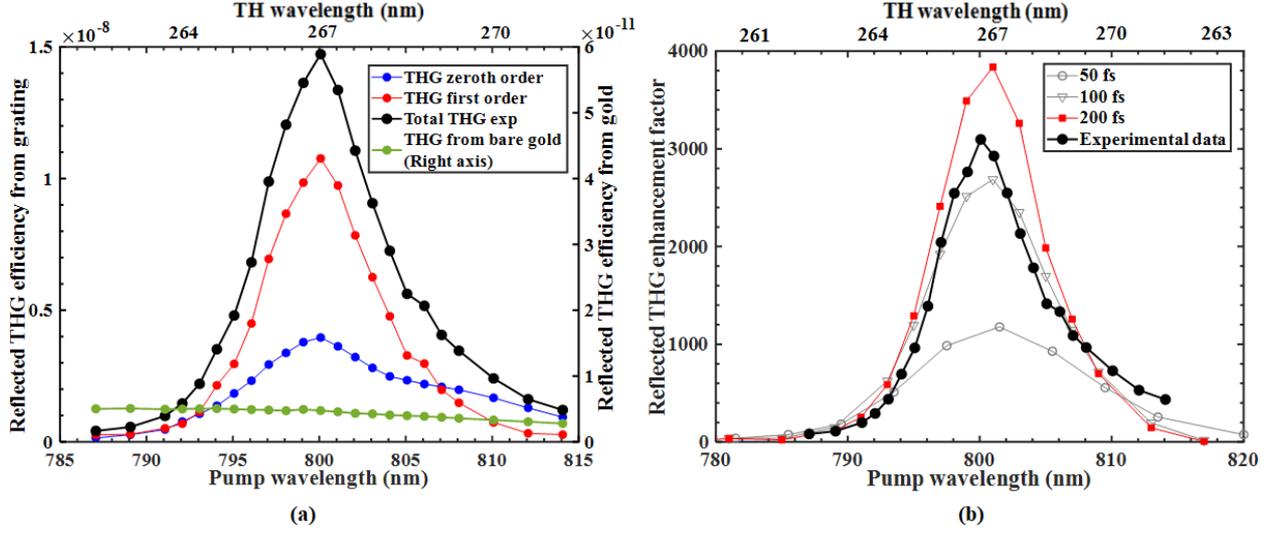

**Figure 7:** (a) Measured efficiency of reflected THG as a function of the carrier wavelength at 16º (blue curve) and 44º (red curve), and total (black curve) normalized to the reflected TH efficiency of a bare gold mirror (green curve at the bottom, right axis); (b) Measured (black curve, as in 7a) and simulated reflected THG efficiency from the grating with pulses of different durations. Both diffraction orders are collected simultaneously.

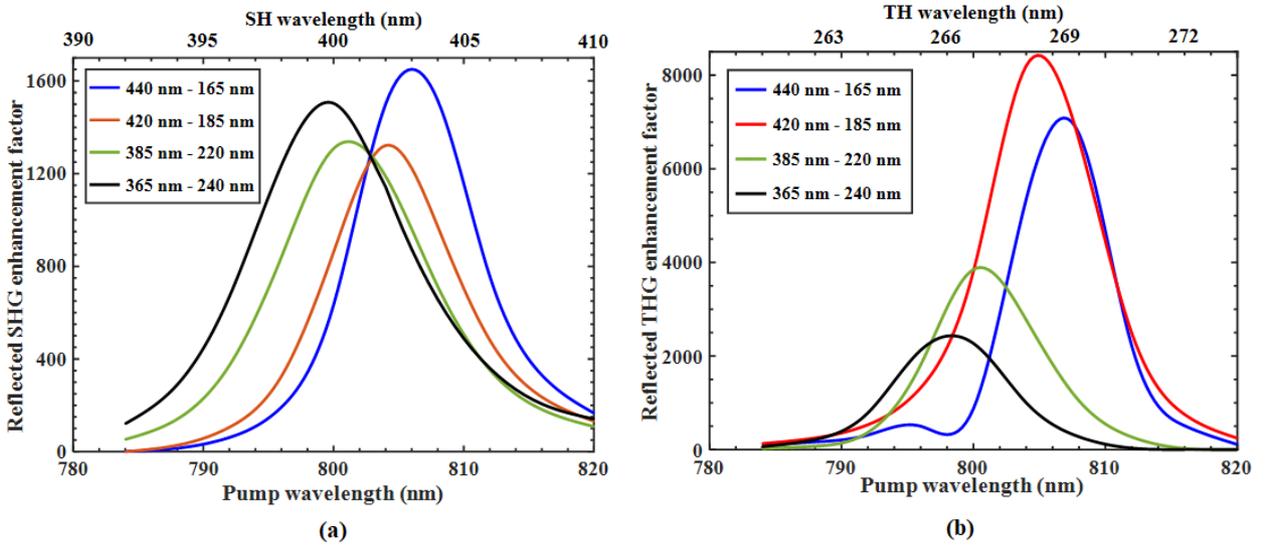

**Figure 8:** Predicted SHG (a) and THG (b) enhancement factors relative to a flat gold mirror as a function of channel and ridge widths. The first number indicates channel width in nanometers. The second number indicates the width of the ridge, also in nanometers. As an example, the red curves denoted by 420nm-185nm represents a gold grating having channel width 420nm, and ridge width 185nm. The periodicity of all gratings remains the same. Aside from a slight redshift of the resonance, the figure shows that the THG enhancement factor can easily be increased by nearly a factor of two without significant geometrical modifications. SHG efficiencies are not affected appreciably because they depend mostly on surface charge densities, currents, and spatial derivatives of the field, rather than field localization.

**Summary**

We have reported SH and TH emissions from a gold grating at different wavelengths and compared to the efficiencies triggered by a flat gold mirror. Using our theoretical model, we find unprecedented agreement between our simulations and experimental observations of both simultaneous second and third harmonic generation from a gold grating at visible and UV wavelengths. This comparison shows dramatic enhancement of SHG and THG conversion efficiencies by more than three orders of magnitude. We have also shown predictions that simple geometrical rearrangements can improve THG efficiency, leaving open the possibility that optimizations can significantly increase TH emissions in the UV range. Geometry affects the SH signal only marginally, since it depends mostly on surface charge and current densities, and spatial derivatives of the fields, which do not change appreciably for the cases we have examined. The enhancement holds at different wavelengths, although field penetration depth can modify the absolute values of conversion efficiencies. The usual emission from bare metal layers shows a characteristic angular dependence that favors maximum SH emission at large angles, and maximum THG at normal incidence [26]. However, photonic devices often require small angle applications. By using a suitable grating with the appropriate periodicity, one may thus achieve SH and TH emissions by the metal grating at almost any desired angle, taking full advantage of diffraction orders, with enhancement factors that are larger than three orders of magnitude with respect to the emission by the bare surface. Combining the various aspects that we have discussed, including pulse duration and geometrical considerations, one may surmise that metals should not be discounted as suitable frequency converters, while more optimized, complex topologies may catalyze further improvements in conversion efficiencies. From a more practical point of view, the converted signals originating in all diffraction orders may be harnessed using properly designed micro-optic lenses placed on top of the grating.

**ACKNOWLEDGEMENTS**

LRS, JT, and CC acknowledge Spanish Agencia Estatal de Investigación (project no. PID2019-105089GB-I00/AEI/10.130397501100011033) and US Army Research Laboratory Cooperative Agreement Nº W911NF1920279 issued by US ARMY ACC-APG-RTP.